\documentclass[12pt]{article}
\usepackage{epsf,amsmath}
\begin{document}

\title{Time and setting dependent instrument parameters
and proofs of Bell-type inequalities}

\author{Karl Hess$^1$ and Walter Philipp$^2$}

\date{$^1$ Beckman Institute, Department of Electrical Engineering
and Department of Physics,University of Illinois, Urbana, Il 61801
\\ $^{2}$ Beckman Institute, Department of Statistics and Department of
Mathematics, University of Illinois,Urbana, Il 61801 \\ }
\maketitle

\begin{abstract}

We show that all proofs of Bell-type inequalities, as discussed in
Bell's well known book and as claimed to be relevant to
Einstein-Podolsky-Rosen type experiments, come to a halt when
Einstein-local time and setting dependent instrument parameters
are included.

\end{abstract}

We have criticized in previous work \cite{hp}-\cite{hpp3} the
various proofs of Bell-type inequalities as given in
\cite{bellbook}, \cite{peres}. Other authors have put forward
related criticism of which we quote only some of the latest
publications \cite{acc}-\cite{acc1}. A very interesting discussion
has developed during the last year \cite{recmer}-\cite{hpm2}. In
the present paper we concatenate our arguments into what we call
``row" and ``column" arguments. These arguments contain reasoning
that is essential to any Bell-type proof. As we show, these
arguments can not be completed when setting and time dependent
instrument parameters are involved. This conclusion is obtained
independently of our previous paper where we derive the quantum
result \cite{hpp2}. We first review the parameter space introduced
by Bell and our extension of this parameter space. We use a
notation that is close to our previous papers
\cite{hp}-\cite{hpp3}. However, for reasons of clarity we
capitalize here all random variables and use the lower case for
the values these random variables can assume.

Bell's \cite{bellbook} parameter random variables are essentially
given by the functions ${A_{\bf a}}(\Lambda) = \pm1, {B_{\bf
b}}(\Lambda) = \pm1$ that are related to the possible outcomes of
spin measurements, with $\Lambda$ being a parameter random
variable that is related to information carried by the correlated
particle pair that is sent out from a common source to two
stations $S_1$ and $S_2$. We assume with Bell and others that the
way Einstein-Podolsky-Rosen (EPR)- experiments are performed
guarantees that $\Lambda$ is independent of the instrument
settings ${\bf a}, {\bf b}$. The form of the experiment was
proposed originally by Bohm and Hiley \cite{bohm}.

We extend this parameter space \cite{hp}-\cite{hpp3} by adding
setting and time dependent instrument parameter random variables
$\Lambda_{{\bf a}, t}^*$ specific to station $S_1$ and
$\Lambda_{{\bf b}, t}^{**}$ to station $S_2$. These variables may
be stochastically independent of $\Lambda$. As an illustration,
these variables $\Lambda_{{\bf a}, t}^*$, $\Lambda_{{\bf b},
t}^{**}$ can be thought of being generated by two computers with
equal internal computer clock time but otherwise entirely
independent. The variables could be represented by any programs
that evaluate the input of ${\bf a}, t$ etc. We do not claim
knowledge of any mathematical properties of these parameters as
dictated by physics nor do we claim that they must exist in
nature. We can currently not simulate the EPR experiment on such
computers and never have claimed that we can. However, we
postulate that any proof of Bell-type inequalities that claims
relevance to locality questions must pass the test to be able to
include such parameters. These parameters do obey Einstein
locality and, therefore, must be covered by any EPR model that is
constructed like Bell's and has the same purpose.

Such parameters require a setting and time dependent joint
probability distribution
\begin{equation}
{\rho_s}(\Lambda_{{\bf a}, t}^*, \Lambda_{{\bf b}, t}^{**},
\Lambda) \label{sp1}
\end{equation}
The subscript $s$ of $\rho$ indicates the setting dependence. The
setting and time dependent parameter random variables
$\Lambda_{{\bf a}, t}^*, \Lambda_{{\bf b}, t}^{**}$ may be
stochastically independent of $\Lambda$. Bell and all his
followers exclude such parameters since they all have only one
probability distribution $\rho(\Lambda)$ that does not depend on
any setting. The instrument parameters that they do include are
assumed to be conditionally independent given $\Lambda$ i.e. if
instrument parameters are included, then they have a product
distribution (see \cite{bellbook} pp36):
\begin{equation}
{\rho_s}(\Lambda_{{\bf a}, t}^*, \Lambda_{{\bf b}, t}^{**}|
\Lambda) = {\rho_1}(\Lambda_{{\bf a},
t}^*|\Lambda)\cdot{\rho_2}(\Lambda_{{\bf b}, t}^{**}|\Lambda)
\label{sp2}
\end{equation}
Because we consider time correlations, Eq.(\ref{sp2}) cannot hold
for our setting and time dependent instrument parameters. Thus
Bell-type proofs exclude a large set of joint probability
distributions from their considerations. In the remainder of the
paper, we deal with the question whether Bell-type proofs can be
reformulated to include such setting and time dependent random
variables. The answer will be in the negative.

Why were such instrument variables not considered? In our opinion
because there was the belief that the requirement
\begin{equation}
A_{\bf a} =  -B_{\bf a}\label{sp3}
\end{equation}
could not be fulfilled if correlations other than those given by
$\Lambda$ would be invoked. In fact, Bell himself writes on p 38
of \cite{bellbook} that for the case that Eq.(\ref{sp3}) holds,
``the possibility of the results depending on hidden variables in
the instruments can be excluded from the beginning." Bell clearly
did not consider the possibility of time correlations such as
\begin{equation}
A_{\bf a}(\Lambda, \Lambda_{{\bf a}, t}^*, {t_{meas.}})  =
-B_{\bf a}(\Lambda, \Lambda_{{\bf a}, t}^{**}, {t_{meas.}})
\label{sp4}
\end{equation}
where the measurement times $t_{meas.}$ in the two stations are,
for the same correlated pair, either the same (as indicated above)
or at least linearly related and therefore can lead to
Eq.(\ref{sp4}) which is equivalent to Eq.(\ref{sp3})  .

Can Bell-type proofs be saved by some reasoning in the extended
parameter space? We show below, that all Bell type proofs contain
what we call ``row" and ``column" arguments that can not be
completed for the extended parameter space. We first consider a
prototype of Bell's original proof arranged according to two types
of reasoning (row and column):

\begin{itemize}

\item[(b1)]The row argument of Bell:

For $x, y, z = {\pm} 1$ we have
\begin{equation}
|xz -yz| = |x - y| = 1 - xy \label{spa1}
\end{equation}
Substituting $x = {A_{\bf b}}({\Lambda})$, $y = {A_{\bf
c}}({\Lambda})$ and $z = {A_{\bf a}}({\Lambda})$ gives
\begin{equation}
|{A_{\bf a}}({\Lambda}){A_{\bf b}}({\Lambda}) - {A_{\bf
a}}({\Lambda}){A_{\bf c}}({\Lambda})| = 1 - {A_{\bf
b}}({\Lambda}){A_{\bf c}}({\Lambda})\label{sp5}
\end{equation}

\item[(b2)]The column argument of Bell: using the inequality

\begin{equation}
{|\int f|} \leq {\int |f|} \label{sp6}
\end{equation}
and the assumption that $\rho$ is a probability density
independent of the settings, we obtain
\begin{equation}
|{\int{({A_{\bf a}}({\Lambda}){A_{\bf b}}({\Lambda}) - {A_{\bf
a}}({\Lambda}){A_{\bf c}}({\Lambda}))}{\rho}(\Lambda) d{\Lambda}}|
\leq {1 - \int{{A_{\bf b}}({\Lambda}){A_{\bf
c}}({\Lambda})}{\rho}(\Lambda) d{\Lambda}}\label{sp7}
\end{equation}
In view of the definition of the expectation value for the spin
pair correlation $E(A_{\bf a}{\cdot}B_{\bf b})$, this yields
Bell's inequality.

\end{itemize}

Does Bell's proof go forward when setting and time dependent
instrument parameter random variables $\Lambda_{{\bf a}, t}^*,
\Lambda_{{\bf b}, t}^{**}$ are included?

\begin{itemize}

\item[(hp1)] The row argument with setting and time dependent
instrument parameters leads to:

\begin{equation}
|{A_{{\bf a},t_1}}({...}){A_{{\bf b},t_1}}({...}) - {A_{{\bf
a},t_2}}({...}){A_{{\bf c},t_2}}({...})| = ?\label{sp8}
\end{equation}
one can see immediately that inclusion of a time index that is
related to the actual time of the measurement does not permit
completion of Bell's reasoning as used in Eq.(\ref{sp5}).

\item[(hp2)] The column argument with setting and time dependent
instrument parameters:

Eq.(\ref{sp7}) is based on the row argument expressed by
Eq.(\ref{sp5}). Because Eq.(\ref{sp5}) can not be derived when our
instrument parameters are included, the column argument can not be
completed as well. In addition, the integration performed in
Eqs.(\ref{sp6}) and (\ref{sp7}) involves now three different joint
probability distributions and corresponding probability measures
$\mu$. Consequently, a fortiori, the integration does not lead to
the inequality expressed in Eq.(\ref{sp7}) and to Bell's
inequality as can be seen by considering the following integrals
on the left side of Eqs.(\ref{sp7}):
\begin{equation}
|\int{A_{{\bf a},t_1}}({...}){A_{{\bf
b},t_1}}({...})d{\mu}({\Lambda}, {\Lambda}_{{\bf a},t}^{*},
{\Lambda}_{{\bf b},t}^{**}) - \int{A_{{\bf a},t_2}}({...}){A_{{\bf
c},t_2}}({...}) d{\mu}({\Lambda}, {\Lambda}_{{\bf a},t}^{*},
{\Lambda}_{\bf c}^{**})|  = ?\label{sp9}
\end{equation}

\end{itemize}

Although this shows already clearly why Bell-type proofs do not go
forward with time and setting dependent instrument parameters, we
will repeat our reasoning more extensively by discussing tables of
possible outcomes for the random variables which is a frequently
used argument in proofs of Bell's inequality. We start again using
only the parameter space of Bell.

\begin{itemize}

\item[(bt1)]The row and column argument of Bell for tables of
possible outcomes.

It is common practice \cite{peres} to form and discuss tables of
possible outcomes that, invariably, involve in each row a certain
sum of terms that we denote by $\Delta$:
\begin{equation}
\Delta = {A_{\bf a}}(\Lambda){B_{\bf c}}(\Lambda) -{A_{\bf
a}}(\Lambda){B_{\bf b}}(\Lambda)-{A_{\bf d}}(\Lambda){B_{\bf
b}}(\Lambda) -{A_{\bf d}}(\Lambda){B_{\bf c}}(\Lambda)
\label{sp10}
\end{equation}
At this point the following statistical argument is usually
invoked in one form or another. If one considers possible outcomes
$\lambda^i$ that the random variable $\Lambda$ may assume, then by
the strong law of large numbers the values ${\lambda}^i$ will
appear approximately the same number of times for each of the
setting pairs since these occur with the same probability. The
reason for this argument is that the source parameter $\Lambda$
does not depend on settings. However, this argument works only if
the cardinality of the set $\{{\lambda^i}\}$ of values that
$\Lambda$ can assume is much smaller than the number of
experiments performed. If all this is fulfilled, then one can
reorder the possible outcomes of $\Lambda$ in a thought experiment
such that one has rows of four terms with the same value
$\lambda^i$ for each element of any given row:
\begin{equation}
{{\Delta}^i} = {A_{\bf a}}(\lambda^i){B_{\bf c}}(\lambda^i)
-{A_{\bf a}}(\lambda^i){B_{\bf b}}(\lambda^i)-{A_{\bf
d}}(\lambda^i){B_{\bf b}}(\lambda^i) -{A_{\bf
d}}(\lambda^i){B_{\bf c}}(\lambda^i) = \pm2 \label{sp11}
\end{equation}

Note that this possibility of reordering makes it unnecessary, at
least in principle, to involve counterfactual arguments i.e.
arguments of what would have happened if a different setting were
chosen. One simply argues statistically that $\Lambda$ will assume
the same values no matter what the setting is and therefore one
can reorder to obtain the table shown immediately below. We will
see, however, that no such reordering is obvious for the extended
parameter space and that one does need counterfactual reasoning to
proceed with Bell-type arguments in that extended space. We will
also see that mere counterfactual reasoning that still might be
admissible is not sufficient to complete the Bell-type proofs.
Because the above equation is true for each row, no separate
column argument is needed to derive Bell-type inequalities. One
can write out the following table and state: within Bell's
original assumptions, the following table is ``sampled" by the
experimental procedure of EPR experiments:

\begin{equation}
\left[ \begin{array}{c}
      {\lambda^1}\\
      {\lambda^2}\\
      \vdots\\
      {\lambda^i}\\
      \vdots\\
      {\lambda^M} \end{array} \right]
\left[ \begin{array}{cccc}
      +{A_{\bf a}^1}{B_{\bf c}^1} & -{A_{\bf a}^1}{B_{\bf b}^1} & -{A_{\bf d}^1}{B_{\bf
      b}^1} & -{A_{\bf d}^1}{B_{\bf c}^1}\\
      +{A_{\bf a}^2}{B_{\bf c}^2} & -{A_{\bf a}^2}{B_{\bf b}^2} & -{A_{\bf d}^2}{B_{\bf
      b}^2} & -{A_{\bf d}^2}{B_{\bf c}^2}\\
      \vdots & \cdots & \cdots & \vdots\\
      +{A_{\bf a}^i}{B_{\bf c}^i} & -{A_{\bf a}^i}{B_{\bf b}^i} & -{A_{\bf d}^i}{B_{\bf
      b}^i} & -{A_{\bf d}^i}{B_{\bf c}^i}\\
      \vdots & \cdots & \cdots & \vdots\\
      +{A_{\bf a}^M}{B_{\bf c}^M} & -{A_{\bf a}^M}{B_{\bf b}^M} & -{A_{\bf d}^M}{B_{\bf
      b}^M} & -{A_{\bf d}^M}{B_{\bf c}^M} \end{array} \right]
 =
\left[ \begin{array}{c}
      \pm2\\
      \pm2\\
      \vdots\\
      \pm2\\
      \vdots\\
      \pm2 \end{array} \right] \label{tb2}
\end{equation}

Although the measurements of the various terms can only be made in
sequence, at the end of the day one would have accumulated this
table and obtained it by reordering provided that the element of
physical reality that corresponds to a given ${\lambda}^i$ could
be somehow made visible. Of course, there may be some terms left
over as implied by the application of the law of large numbers.
However, the number of such incomplete rows is negligible for
large $M$.

\end{itemize}

With time and setting dependencies permitted we need to invoke the
following table in an attempt to proceed with reasoning similar to
the above:
\begin{equation}
\left[ \begin{array}{cc}
      {\lambda^1} & {t_1}\\
      {\lambda^2} & {t_2}\\
      \vdots & \vdots\\
      {\lambda^i} & {t_i}\\
      \vdots & \vdots\\
      {\lambda^M} & {t_M} \end{array} \right]
\left[ \begin{array}{cccc}
      +{A_{\bf a}}{B_{\bf c}^*} & -{A_{\bf a}}{B_{\bf b}} & -{A_{\bf d}}{B_{\bf
      b}} & -{A_{\bf d}}{B_{\bf c}}\\
      +{A_{\bf a}}{B_{\bf c}} & -{A_{\bf a}}{B_{\bf b}} & -{A_{\bf d}}{B_{\bf
      b}^*} & -{A_{\bf d}}{B_{\bf c}}\\
      \vdots & \cdots & \cdots & \vdots\\
      +{A_{\bf a}}{B_{\bf c}} & -{A_{\bf a}}{B_{\bf b}^*} & -{A_{\bf d}}{B_{\bf
      b}} & -{A_{\bf d}}{B_{\bf c}}\\
      \vdots & \cdots & \cdots & \vdots\\
      +{A_{\bf a}}{B_{\bf c}} & -{A_{\bf a}}{B_{\bf b}} & -{A_{\bf d}}{B_{\bf
      b}} & -{A_{\bf d}}{B_{\bf c}^*} \end{array} \right]
 =
\left[ \begin{array}{c}
      ?\\
      ?\\
      \vdots\\
      ?\\
      \vdots\\
      ? \end{array} \right] \label{tb1}
\end{equation}
Here the asterisk for one term per row is used to indicate that
only one pair of settings is possible at a given time. The
question-mark $?$ on the right hand side replaces now the $\pm2$
for reasons given immediately by discussing row and column
arguments. Before doing so we digress to clarify the meaning of
elements of physical reality that are related to the hidden
parameters.

The EPR argument postulates a reality for the parameters or
information that come with the particles from the source. If
instrument parameters $\Lambda_{{\bf a}, t}^*, \Lambda_{{\bf b},
t}^{**}$ are introduced, then the reality of what is measured or
the information content of what is measured depends now on both
the information from the source and the information from the
instruments in a mixed way. We are not talking about the reality
of the information of the source alone but about that reality as
``seen" through the instruments. That mixed information depends on
the actual macroscopic settings of the instruments ${\bf a}, {\bf
b}$ etc.. As a consequence this mixed reality can not exist for
different settings at the same time. Therefore, all arguments
involving all of Table(\ref{tb1}) are now counterfactual. It is
important then to explain how results from counterfactual
arguments can be compared to experiments. Furthermore, one needs
to deal with the fact that only the fraction of Table(\ref{tb1})
marked by asterisks does correspond to possible values that the
random variables may assume. Only one value may be assumed in a
row, any second value is impossible, much as a coin toss can give
only head or tail but not both at a time. The quantity $\Delta$,
representing a row, is therefore not even a random variable as can
be seen from the textbook definition (P. Halmos): ``a random
variable is a function attached to an experiment - once the
experiment has been performed the value of the function is known."
These facts invalidate all Bell-type proofs known to us as can
immediately be seen from the following.

\begin{itemize}

\item[(hpt1)] The row argument for tables of possible outcomes with
time and setting dependent instrument parameters:

The pair of settings that are chosen can be thought of as picked
by the throw of a tetrahedal die. Therefore at a given time only
one pair can be picked. Adding or counting more of the possible
(for themselves) outcomes of a row amounts to the same as adding
or counting different possible outcomes for tosses of the
tetrahedal die (or of coin tosses as mentioned above) at a given
time. The result of such a procedure is, in general, bound to be
incorrect as can be shown by numerous examples. Therefore, the row
argument does not work and cannot be used to compute the row-sum
of $\pm2$ in Table(\ref{tb1}).

\item[(hpt2)] The column argument for tables of possible outcomes
with time and setting dependent instrument parameters:

Bell-type proofs need to show that the sampling of
Table~(\ref{tb1}) leads to an expectation value of $\Delta$ that
represents essentially Bell-type inequalities. However, $\Delta$
is not what is necessarily sampled by any procedure commensurate
with the experiments as can be seen from the fact that one has
four different integrals for the four expectation values of the
spin pair correlation corresponding to the four columns e.g.
\begin{equation}
E(A_{\bf a}{\cdot}B_{\bf b}) = \int A_{\bf a}B_{\bf
b}{\rho_s}(\Lambda_{{\bf a}, t}^*, \Lambda_{{\bf b}, t}^{**},
\Lambda) d{\Lambda_{{\bf a}, t}^* d\Lambda_{{\bf b}, t}^{**}
d\Lambda}\nonumber
\end{equation}
or
\begin{equation}
E(A_{\bf a}{\cdot}B_{\bf c}) = \int A_{\bf a}B_{\bf
c}{\rho_s}(\Lambda_{{\bf a}, t}^*, \Lambda_{{\bf c}, t}^{**},
\Lambda) d{\Lambda_{{\bf a}, t}^* d\Lambda_{{\bf c}, t}^{**}
d\Lambda}\nonumber
\end{equation}
and similar with other settings that appear in the Bell-type
inequalities. Only if the joint distributions $\rho_s$ are
independent of the setting can the Bell-type inequalities be
obtained the way they are usually proven.

\end{itemize}

We have thus excluded the main arguments that can be used to prove
the Bell inequalities when time and setting dependent parameter
random variables are involved i.e. the row and column arguments
and the argument involving reordering. Below we attempt to shine
some additional light on the failure of these arguments using
slightly different viewpoints. It is important to be clear about
the following fact. When time dependencies are included, the
Bell-type arguments become necessarily counterfactual i.e. they
involve considerations that can not be proven by experiment. This
follows simply from the fact that any reasoning that involves
considerations of different settings at the same time does not
correspond to an actual experiment because at a given time only
one setting of a macroscopic instrument is possible. The
discussions above show also clearly that the technical expression
``counterfactual" is, by itself, not sufficient to describe the
extent of its meaning for a proof involving mathematical logic and
deductions thereof. It definitely is still permissible to
consider, as Einstein did, the possibility that a different
setting could have been used. Recall Einstein's original argument:
if setting $\bf a$ is used in station $S_1$ and $A_{\bf a} = +1$
then one can predict with probability one the result $B_{\bf a} =
-1$ in station $S_2$. Had instead setting $\bf b$ been used in
$S_1$ and the result was $A_{\bf b} = +1$, then one can predict
with probability one the result $B_{\bf b} = -1$ in $S_2$. We can
not see anything wrong in the reasoning of this argument. However,
if anyone starts adding or subtracting outcomes for settings that
could have been chosen, then the result of such addition may not
have any meaning and can be completely nonsensical. A simple
example makes this clear. Consider ordering a meal from the menu
in a restaurant. After the meal, the owner of the restaurant adds
possible other choices that you could have made and presents the
bill for all choices. Note, however, that in the standard
text-book of Peres \cite{peres} such arguments occur in abundance.
For instance on p 167 of \cite{peres} the author reports on
``people who ponder over a menu in a restaurant"-``with no
apparent ill effects", while on p 163 of \cite{peres} the author
concedes with regard to his Eq.(6.24) that not all tests can be
performed simultaneously. That fact does not prevent the author
from totaling up the ``results" of such non-performed experiments
in his Eqs.(6.24) and (6.28). We therefore should name such
situations as described above not just counterfactual but, rather
``counter-syntaxial" \cite{stout} indicating that the
counterfactual argument is compounded by the use of mathematical
operations such as counting, adding, subtracting, etc. in a way
that violates well established protocol. While counterfactual
reasoning still may permit to continue with mathematical logic,
counter-syntaxial reasoning invalidates any proof. The row
argument in Bell-type proofs is counter-syntaxial and therefore
mathematically inadmissible. What remains, therefore, is the
column argument. In other words, Bell-proof supporters must show
that the counterfactual Table~(\ref{tb1}) is still sampled in its
entirety with high statistical accuracy because of the fact that a
large fraction of the columns is sampled. However, we have shown
above that this argument does not lead to Bell-type inequalities
when setting and time dependent instrument parameters are involved
because these imply the existence of setting dependent joint
probability distributions. We have seen, however, in many
discussions with colleagues, that these facts are not recognized
because of a circular argument that can be summarized as follows.

It is often simply being assumed that what is sampled (e.g. in a
Monte Carlo integration sense) by the procedure of random choice
of setting, time and $\Lambda$ is just Table~(\ref{tb1}), even if
time correlations are involved. Others assume that if one would
reorder the measured results one would invariably end up with a
table equivalent to the full Table~(\ref{tb1}) ignoring the fact
that the measurements are all at different times and that the
reordering procedure would require extensive mathematical proof.
Since this Table~(\ref{tb1}) leads immediately (by trivial
algebra) to the Bell inequalities, the Bell inequalities are then
assumed to be proven. This is clearly a logical circle. The task
is to prove that any sampling procedure commensurate with the
experiment or any such reordering procedure indeed gives a table
equivalent to Table~(\ref{tb1}). That this is not possible by
general mathematical methods can be seen from the following facts.
Any sampling procedure that is commensurate with the experiments
samples functions $A_{\bf a}, B_{\bf b}$ on a space that includes
the respective settings, $\Lambda$ and $\Lambda_{{\bf a}, t}^*$,
$\Lambda_{{\bf b}, t}^{**}$ at time of measurement $t_{meas.}$.
One can say that one samples the product $A_{\bf a}(\Lambda,
\Lambda_{{\bf a}, t}^*,  t_{meas.}){\cdot}B_{\bf b}(\Lambda,
\Lambda_{{\bf b}, t}^{**}, t_{meas.})$ by choosing randomly values
of ${\bf a},{\bf b}, \Lambda, \Lambda_{{\bf a}, t}^*,
\Lambda_{{\bf b}, t}^{**}$ and $t_{meas.}$. Involving reasonable
conditions for the parameters, this is indeed true and the result
relates directly to the expectation value $E(A_{\bf
a}{\cdot}B_{\bf b})$ of the spin pair correlation. To prove the
Bell inequalities, however, one wants to sample $\Delta$ and not
just the $A.B$ products. However, $\Delta$ is not randomly
sampled, because each value of $\Delta$ involves four $A{\cdot}B$
products taken at the same time and with the same $\Lambda$ for
four different settings while only one term of any row is sampled
by the measurement procedure. Think of the sampling as a Monte
Carlo integration procedure. To sample $\Delta$, such an
integration scheme must include random choices in any given row
including two, three or four elements at the same time with the
same value that $\Lambda$ may have assumed and for different
settings. However, only one such element is sampled per row. This
problem is compounded by the fact that each column contains
different stochastic variables that are setting dependent and
follow different joint distributions. From this one can see that
$\Delta$ is not sampled. Nor can we see any procedure of
reordering that would map any possible outcomes that correspond to
actual measurements onto Table~(\ref{tb1}). This brings home the
importance of the definition of a random variable and the
importance to adhere to that definition when probability theory is
used. According to this definition, $\Delta$ is not a random
variable and Bell-type proofs are therefore mathematically
questionable in general. They become invalid when time and setting
dependent instrument parameters are involved.

One can now ask the question whether there can be any other proof
of the Bell-type which can deal with setting and time dependent
instrument parameters. We have attempted to answer this question
in reference \cite{hpp2} and we point the reader to this
reference. There we show, that for any random sequence of setting
pairs (that appear in the Bell inequalities) there exists a
distribution of local setting and time dependent instrument
parameters that leads to the quantum result for the expectation
value $E(A_{\bf a}{\cdot}B_{\bf b})$ of the spin pair correlation.
This model shows therefore that one can construct joint
probability distributions that do not lead to Bell type
inequalities and that $\Delta$ is not necessarily sampled by the
experimental procedure. Therefore, assuming the validity of the
results given in \cite{hpp2} , no Bell-type proof can be given
because such proofs always need to show that $\Delta$ is sampled
by the experimental procedure if the parameters involved obey
Einstein locality. The discussion between Einstein and Bohr would
then be undecidable by a probability model of the type that Bell
suggested.

Acknowledgement: The work was supported by the Office of Naval
Research N00014-98-1-0604.

\end{document}